\documentclass[sigconf,nonacm]{acmart}
\AtBeginDocument{%
  }

\setcopyright{none}
\begin{document}

\title{Multimodal Fusion And Sparse Attention-based Alignment Model for Long Sequential Recommendation}

\author{Yongrui Fu}
\orcid{0009-0001-0309-8264}
\affiliation{%
  \institution{Fudan University}
  \city{Shanghai}
  \country{China}}
\email{23210240154@m.fudan.edu.cn}

\author{Jian Liu}
\authornote{Corresponding author.}
\affiliation{%
  \institution{Baidu, Inc}
  \city{Beijing}
  \country{China}}
\email{liujian21@baidu.com}

\author{Tao Li}
\affiliation{%
  \institution{Baidu, Inc}
  \city{Beijing}
  \country{China}}
\email{litao@baidu.com}

\author{Zonggang Wu}
\affiliation{%
  \institution{Baidu, Inc}
  \city{Beijing}
  \country{China}}
\email{wuzonggang@baidu.com}

\author{Shouke Qin}
\affiliation{%
  \institution{Baidu, Inc}
  \city{Beijing}
  \country{China}}
\email{qinshouke@baidu.com}

\author{Hanmeng Liu}
\affiliation{%
  \institution{Baidu, Inc}
  \city{Beijing}
  \country{China}}
\email{liuhanmeng@baidu.com}

\renewcommand{\shortauthors}{}

\begin{abstract}
Recent advances in multimodal recommendation enable richer item understanding, while modeling users' multi-scale interests across temporal horizons has attracted growing attention. However, effectively exploiting multimodal item sequences and mining multi-grained user interests to substantially bridge the gap between content comprehension and recommendation remain challenging. To address these issues, we propose \textbf{MUFASA}—a \textbf{MU}ltimodal \textbf{F}usion \textbf{A}nd \textbf{S}parse Attention-based \textbf{A}lignment model for long sequential recommendation. Our model comprises two core components. First, the Multimodal Fusion Layer (MFL) leverages item titles as a cross-genre semantic anchor and is trained with a joint objective of four tailored losses that promote: (i) cross-genre semantic alignment, (ii)alignment to the collaborative space for recommendation, (iii)preserving the similarity structure defined by titles and preventing modality representation collapse, and(iv) distributional regularization of the fusion space. This yields high-quality fused item representations for further preference alignment. Second, the Sparse Attention–guided Alignment Layer(SAL) scales to long user-behavior sequences via a multi-granularity sparse attention mechanism, which incorporates windowed attention, block-level attention, and selective attention, to capture user interests hierarchically and across temporal horizons. SAL explicitly models both the evolution of coherent interest blocks and fine-grained intra-block variations, producing robust user and item representations. Extensive experiments on real-world benchmarks show that MUFASA consistently surpasses state-of-the-art baselines. Moreover, online A/B tests demonstrate significant gains in production, confirming MUFASA's effectiveness in leveraging multimodal cues and accurately capturing diverse user preferences.

\end{abstract}




\maketitle
\section{Introduction}
Recent advances in recommender systems have demonstrated that collaborative filtering (CF) methods~\cite{he2016ups, chen2017sampling, ding2020simplify}, which rely on historical user–item interactions, suffer from the cold-start challenges for new items due to lacking interaction data.  Content-based (CB) approaches~\cite{van2013deep} focus on extracting item content features and can partially alleviate cold-start issues; however, due to the “semantic gap” between content features and users’ behavioral interests, they struggle to deliver high-quality personalized recommendations~\cite{barkan2019cb2cf}. Consequently, both academia and industry have explored bridge models that map item information into the CF latent space, aiming to satisfy the dual demands of cold-start handling and accurate personalization. Simultaneously, long-sequence recommendation techniques~\cite{yu2019adaptive}, by deeply modeling the sequential relationships in user behavior histories, capture complex interest structures and both short-term and long-term preferences. These methods have achieved remarkable success in large-scale applications such as e-commerce and short-video platforms.
\begin{figure}[!t]
\centering
\includegraphics[width=3in]{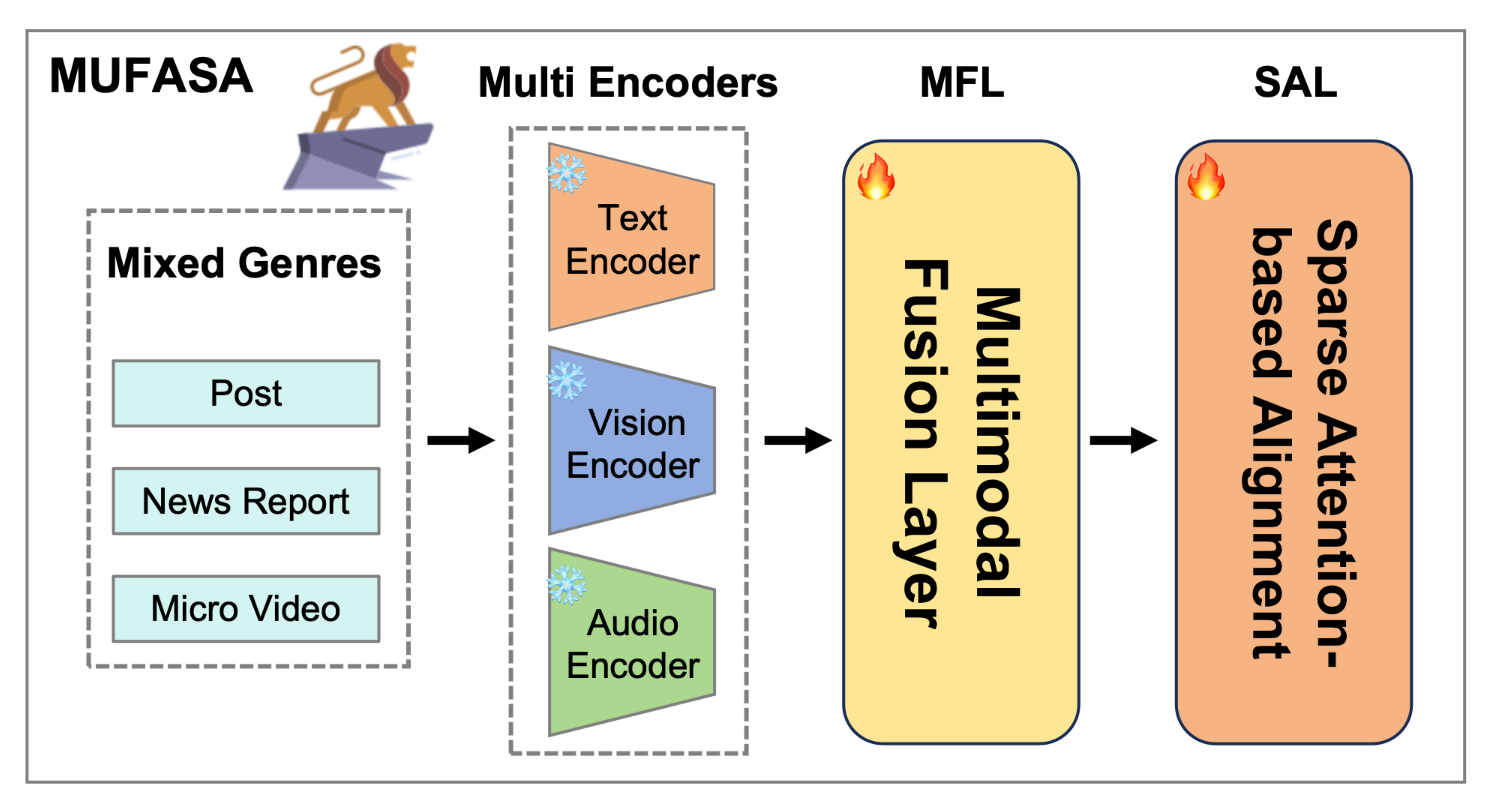}
\Description{The overall architecture of MUFASA.} 
\caption{The overall architecture of MUFASA.}
\label{fig:fig_2}
\end{figure}
Multimodal fusion techniques which combine ID, text, and visual features are widely adopted to enrich item representations and bridge gaps between multimodal content and recommendation targets. For instance, LEARN~\cite{jia2025learn} replaces traditional ID embeddings with LLM-generated text features, improving cold-start performance. Similarly, MLLM-MSR~\cite{ye2025harnessing} integrates multimodal summaries via prompt engineering, enhancing accuracy and interpretability in sparse data environment. However, in industrial mixed-genre applications such as post, news report and micro video recommendations, the increased diversity of modalities and mixed genres presents additional challenges. Effectively managing conflicts and complementarities among these modalities and genres becomes even more critical, and currently, no unified method exists that can efficiently fuse all types of modality embeddings and align different  genre content.

After obtaining fused vectors,  further mining user preferences is essential to bridge the gap between content-driven priors and recommendation-driven semantics~\cite{hidasi2015session}. Predominant approaches~\cite{kang2018self,sun2019bert4rec} leverage full attention mechanisms for preference modeling, yet these exhibit notable limitations: susceptibility to noise interference and prohibitive computational overhead. While sparse attention variants~\cite{beltagy2020longformer,zaheer2020big,lin2025iterative} offer partial mitigation, their foundational limitation persists: treating individual items as atomic attention units induces excessive sparsity and redundancy. This granularity constraint obstructs dynamic interest block detection in behavioral sequences. Prior research~\cite{yuan2025native} has noted that attention in long sequences is inherently sparse, especially in recommendations. Using single items as the minimal attention granularity exacerbates sparsity and introduces additional noise, hindering effective learning of long-sequence user preferences.

Industrial environments amplify these challenges as exponentially accumulating user interaction histories encapsulate multi-scale preferences. Critically, these histories manifest as coherent interest blocks: contiguous interaction subsequences reflecting thematic preference domains. For instance, sequential engagements with athletics content followed by sneaker-related videos constitute a singular sports interest block. Such blocks provide superior interest representations compared to fragmented item-level analysis. Furthermore, natural limits in user attention create uneven focus both between and within these blocks. Consequently, two imperatives emerge: the automated identification of semantically cohesive blocks within long interaction sequences, and the fine-grained intra-block modeling to capture nuanced user interest preferences for personalized recommendation.

To address these challenges, we propose \textbf{MUFASA}—a \textbf{MU}ltimodal \textbf{F}usion \textbf{A}nd \textbf{S}parse Attention-based \textbf{A}lignment model for long-sequence recommendation.

Multimodal content across diverse genres exhibits inherent conflicts such as video-subtitle inconsistency and complementarities such as audio-title synergy, while semantic mismatches persist across posts, news reports, and micro videos. To address these challenges, we propose an innovative approach that utilizes title text embeddings as semantic anchors and introduces a \textbf{M}ultimodal \textbf{F}usion \textbf{L}ayer(\textbf{MFL}). This layer optimizes a four-fold joint objective. First, \textbf{Title-Guided Contrastive Learning} aligns heterogeneous modality vectors across diverse genres by enforcing the correspondence between each item’s fusion features and the semantic space defined by its title. For example, consider the event NBA Finals Game 6 between the Clippers and the Warriors. A social post with a courtside snapshot and brief caption, a news report summarizing the key scoring runs, and a micro video showing the decisive moments each uses its own title as a semantic anchor. By pulling their fused representations toward the corresponding title embeddings, items about the same event are placed in a shared neighborhood of the title space, achieving cross-genre semantic alignment. Second, \textbf{Collaborative Filtering Embedding-Guided Fusion Learning} initially bridges the gap between content representations and recommendation domain by projecting multimodal fused vectors into the collaborative filtering latent space. Third, \textbf{Title-Guided Consistency Constraint} aligns fused embeddings with the similarity structure implied by titles, limiting CF overfitting and preventing multimodal collapse. Fourth, \textbf{Fusion Embedding-Centric Contrastive Learning} further improves cross-sample discriminability and robustness to noise within the unified embedding space.

These specialized losses resolve semantic conflicts across modalities and genres, enabling collaborative cross-modal representation while preserving semantic richness and recommendation suitability. Consequently, generated user/item embeddings achieve high discriminability and semantic coherence, establishing a robust foundation for downstream preference alignment.

For the \textbf{S}parse \textbf{A}ttention-based A\textbf{L}ignment Layer(\textbf{SAL}), we address the inherent sparsity of attention in long sequences through a hierarchical design that captures short-term, long-term, and core preferences at different levels of granularity. To model recent interactions, we employ \textbf{windowed attention}, which effectively captures short-term preferences while circumventing the quadratic computational complexity associated with full attention mechanisms, thereby significantly improving scalability. For modeling overall historical interactions, we integrate \textbf{block-level attention} with \textbf{selective attention} to capture long-term and core preferences. In this layer, multiple consecutive interactions are aggregated into interest blocks that represent diverse domains of user interests, such as gastronomy, gaming, history, or travel, thereby avoiding the redundancy and bias that may arise from considering individual items in isolation. For those blocks that receive heightened user attention, selective attention is employed to learn fine-grained preference distinctions within each block, such as differentiating among various cities within the travel domain or among different games within the gaming category. By integrating these three complementary sparse attention mechanisms, our approach enables hierarchical and multi-granularity modeling of user interests, ranging from fine-grained short-term behaviors to coarse long-term structures and nuanced intra-block preferences. This results in the generation of user embeddings that are both highly expressive and robust to noise.

In summary, our main contributions are:

1. We introduce a multimodal fusion layer(MFL) anchored on title embeddings and a four-fold joint optimization strategy, substantially enhancing vector representation capability.

2. We conceptualize interest blocks as the minimal units of user preference and propose a sparse attention-based alignment layer(SAL) that models user interests at multiple granularities via three attention mechanisms, yielding superior user embeddings.

3. We validate our model’s effectiveness on both large-scale real-world micro-video and mixed genre datasets, demonstrating the efficacy of our proposed modules.

\section{Related Work}
\subsection{multimodal recommendation}
With the increasing diversification of information acquisition methods, items across genres in recommender systems frequently contain multimodal data, including text, images, and audio. Early approaches to multimodal recommendation primarily relied on feature-level concatenation~\cite{snoek2005early,srivastava2012multimodal} or decision-level weighting~\cite{rovid2019towards,wei2020decision}. However, these methods suffer from fundamental shortcomings. Both feature concatenation and decision weighting fail to establish effective cross-modal alignment mechanisms, leading to unresolved inconsistencies and amplified semantic ambiguity across modalities and genres. Moreover, such inadequate fusion strategies inherently ignore inter-modality correlations, ultimately compromising recommendation relevance and effectiveness.

To overcome the static fusion limitations of early methods, multimodal attention has become a core solution. Nagrani et al.~\cite{nagrani2021attention} proposed a novel transformer-based architecture utilizing fusion bottlenecks for modality fusion across layers; HAF-VT~\cite{wu2025hierarchical} introduced hierarchical attention fusion by extracting static text and image features via CLIP and then incorporating two layers of attention to capture cross-domain interest transfer. Meanwhile, in response to the semantic gap between heterogeneous modalities, contrastive learning~\cite{he2020momentum, ye2023dream} has emerged as a key technique for aligning multimodal spaces. For instance, SyCoCa~\cite{ma2024sycoca} proposed Symmetrizing Contrastive Captioners, introducing a TG-MIM head to enable bidirectional local interactions from image to text and text to image, thereby enhancing fine-grained multimodal alignment.

Despite significant technological advancements, multimodal recommendation still faces two persistent challenges. First, current contrastive learning frameworks fail to establish genre-aware alignment objectives, achieving only shallow cross-modal correspondence while lacking mechanisms for deep semantic integration. Crucially, they cannot address the heterogeneous semantic misalignment across genres, such as post, news report and micro video. Second, dynamic preference modeling is limited. Users’ multimodal preferences change with different contexts and over time, yet current frameworks struggle to capture these variations in real time.

\subsection{Long Sequential Recommendation}

Long-sequence recommendation aims to capture both users’ short-term and long-term interests by modeling their historical behavior sequences~\cite{xu2021long, qiu2022contrastive}. Early methods such as FPMC~\cite{chen2018survey} and GRU4Rec~\cite{hidasi2015session} relied on Markov chains or recurrent neural networks to model behavioral dependencies via state transitions or hidden state propagation. However, RNN-based models struggle with long-range dependencies beyond roughly 50 steps, fail to explicitly address user interest evolution, such as when a user shifts from browsing sneakers to collecting travel guides.

DIN~\cite{zhou2018deep} introduced the target item attention mechanism, dynamically weighting the importance of historical behaviors and, for the first time, realizing the concept of “interest activation.” SASRec~\cite{kang2018self} was the first to introduce self-attention into recommendation, capturing associations across the entire item sequence; BERT4Rec~\cite{sun2019bert4rec} leveraged the masked language model to learn context-aware representations via bidirectional attention. However, these models face challenges in long-sequence scenarios, such as excessive computational complexity and inability to capture user interest shifts at varying granularities.

Recent advances such as sparse attention~\cite{beltagy2020longformer, zaheer2020big} have improved efficiency and the modeling of local interests. For example, the ISA model~\cite{lin2025iterative} combines global and local attention to maintain long-range dependencies while reducing computational cost. However, most models still treat individual items as the basic unit, overlooking interest segments and multi-granularity preferences. ComiRec~\cite{cen2020controllable} introduced a multi-interest extraction layer that clusters behavior sequences into several interest capsules using dynamic routing, thus abstracting user interests. Nevertheless, this approach requires a predefined number of capsules and does not clearly define the boundaries of interest segments.

\section{Method}
\begin{figure*}[!t]
\centering
\includegraphics[width=6in]{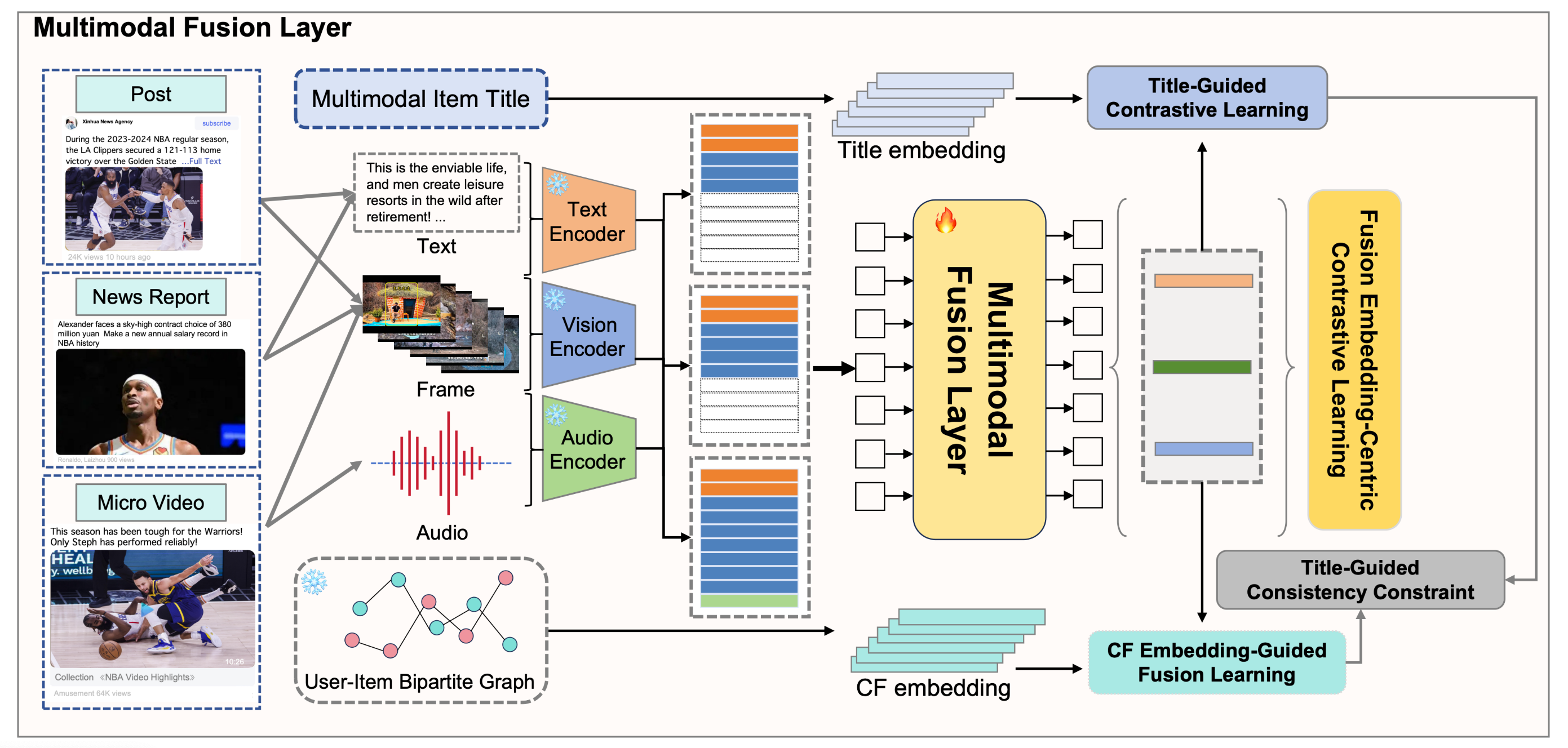}
\Description{Multimodal Fusion Layer (MFL)} 
\caption{Multimodal Fusion Layer (MFL): Modality-specific encoders generate unimodal representations for items across diverse genres. These are concatenated and processed by the MFL to produce a fused multimodal representation, optimized jointly by four losses: Title Embedding Guided Contrastive Learning, Collaborative Filtering Embedding Guided Fusion Learning, Title-Guided Consistency Constraint, and Fusion Embedding Centric Contrastive Learning.}
\label{fig:fig_1}
\end{figure*}

In this section, we will provide detailed design and implementation of our proposed MUFASA. Our MUFASA consists of two modules, in which the MFL is employed to fuse feature representations across heterogeneous modalities and diverse genres, as shown in Fig. \ref{fig:fig_1}; the SAL is used to learn users’ interest preferences across diverse temporal contexts and granularity levels, as shown in Fig. \ref{fig:fig_2}.

\subsection{Multimodal Fusion Layer}
In modern recommender systems, recommended items of multiple genres often contain rich multimodal information. Existing studies, such as Ni et al.~\cite{ni2023content} have pointed out that simply concatenating or adding multimodal representations does not adequately model the complementary and conflicting relationships between different modalities, does not align different genres and fails to achieve effective alignment with the final recommendation semantic space. Consequently, fused vectors exhibit deficient semantics that compromise their direct applicability in sequential recommendation tasks.

To address this issue, we propose a Multimodal Fusion Layer(MFL), situated between the extraction of multimodal item features and user behavior sequential modeling. MFL acts as a bridge that jointly performs multimodal semantic fusion, cross-genre semantic alignment, and preliminary alignment to the recommendation space. Through multi-objective learning, it reconciles complementary and conflicting modalities, unifies heterogeneous genre semantics, and narrows the gap between content understanding and recommendation objectives, providing semantically richer and more consistent inputs to the subsequent Alignment Layer.

MFL consumes pre-trained text, image, audio, and other modality features, each as a fixed-length vector. These vectors are stacked into a tensor of shape $[N,M,d]$, where $N$ denotes the batch size, $M$ represents the number of modalities, and $d$ is the feature dimension. The fusion network outputs a unified vector, which is further refined by a sparse attention–based alignment layer.

Within the MFL, the fused representations are primarily optimized using the following loss functions:

\subsubsection{Title Embedding Guided Contrastive Learning}
To mitigate cross-modal and cross-genre semantic misalignment in multimodal recommendation, we align fused representations with title embeddings through contrastive learning.

Concretely, we utilize the semantic representation of the item title, obtained through a pretrained Text Encoder, as a semantic anchor to guide the alignment of the multimodal fusion vector across diverse genres. For each item, the fusion vector $\mathbf{z}_i$ and its corresponding title embedding $\mathbf{t}_i$ are regarded as a positive sample pair, while the title embeddings of other items within the same mini-batch serve as negative samples. The objective function is formulated as the InfoNCE loss, expressed as follows:

\begin{equation}
\label{loss_title}
\mathcal{L}_{\text{title}} = -\log \frac{\exp(s(\mathbf{z}_i, \mathbf{t}_i) / \tau)}{\sum_{j=1}^N \exp(s(\mathbf{z}_i, \mathbf{t}_j) / \tau)},
\end{equation}

Here, $s(\mathbf{a}, \mathbf{b}) = \mathbf{a}^\top \mathbf{b} / (|\mathbf{a}||\mathbf{b}|)$ denotes the cosine similarity between vectors $\mathbf{a}$ and $\mathbf{b}$, The parameter $\tau$ is a temperature hyperparameter that regulates the concentration level of the distribution and thus controls the contribution of hard negative samples. The variable $N$ represents the batch size, indicating the number of negative samples included. 

By optimizing this objective, the fused representation $\mathbf{z}_i$ is encouraged to be semantically closer to its corresponding title embedding $\mathbf{t}_i$, while being distinguishable from the title embeddings of other items in the batch. This contrastive mechanism concurrently resolves semantic discrepancies and potential conflicts among heterogeneous modalities, while simultaneously enhancing semantic alignment across diverse genres. Overall, the proposed loss guides the mixed-genre item vectors towards a unified and coherent semantic space, thereby enhancing both the quality and interpretability of the multimodal representations.

In practice, this objective imposes stricter requirements on title quality. To ensure reliable supervision, we exclude items with low-quality titles, including missing titles, overly short titles, or semantically vague titles, from contrastive training. We can also leverage LLMs~\cite{ye2025harnessing} to rewrite or enrich titles and to produce higher-quality title embeddings, which strengthens the learning signal and improves training effectiveness.

\subsubsection{Collaborative Filtering Embedding Guided Fusion Learning}
In order to more effectively align the fused representation $\mathbf{z}_i \in \mathbb{R}^d$ with the collaborative filtering semantic space within recommendation systems, we introduce a guidance mechanism based on collaborative filtering embeddings. Specifically, we utilize the collaborative filtering vector $\mathbf{c}_i \in \mathbb{R}^d$, which captures co-occurrence relationships derived from user-item interactions in the user-item bipartite graph, as a supervisory signal for the fused vector. To this end, we minimize the mean squared error loss between the fused representation and the corresponding collaborative filtering embedding, thereby narrowing the semantic gap between multimodal content features and structural co-occurrence patterns. The loss function is formally defined as follows:

\begin{equation}
\label{loss_cf}
\mathcal{L}_{\text{cf}} = \frac{1}{N} \sum_{i=1}^N \|\mathbf{z}_i - \mathbf{c}_i\|^2_2
\end{equation}

In this expression, $N$ denotes the batch size, while the term $\|\cdot\|_2$ represents the Euclidean norm. By minimizing the squared Euclidean distance between the fused vector and the collaborative filtering embedding, this geometric constraint encourages the fused representation $\mathbf{z}_i$ to capture both rich multimodal semantics and collaborative contexts.

Through this collaborative filtering embedding guided learning strategy, the fused vector is refined to align with the graph-based collaborative space. This space structurally encodes collective behavioral contexts derived from interactions such as clicks, likes, and collects. Such alignment bridges multimodal content features with collaborative behavior patterns, thereby enhancing the synergy between content semantics and collaborative signals.

\subsubsection{Title-Guided Consistency Constraint}
However, the $\mathcal{L}_{\text{cf}}$ objective can dominate optimization, rapidly biasing the fused latent space toward the CF manifold and risking multimodal representation collapse. To counter this, we impose a title-guided consistency constraint that keeps the title-based neighborhood structure in the fused space by aligning pairwise similarities of fused embeddings with those from title embeddings:

\begin{equation}
\label{loss_cons}
\mathcal{L}_{\text{cons}} =  \| C(\mathbf{z}_i,\mathbf{z}_j) - C(\mathbf{t}_i,\mathbf{t}_j) \|_2^2 ,
\end{equation}

where $C(\cdot,\cdot)$ represents the cosine similarity function. $\mathbf{z}_i$ and $\mathbf{z}_j$ are fused item embeddings, $\mathbf{t}_i$ and $\mathbf{t}_j$ are the corresponding title embeddings. loss is computed over sampled item pairs $(i, j)$.

Intuitively, $\mathcal{L}_{\text{cf}}$ will drive $z$ toward $c$ too aggressively; $\mathcal{L}_{\text{cons}}$ anchors $z$ to the semantic relational structure of titles so that the fused representation does not drift far from the title-induced distribution, yielding more balanced fusion, stability, and better generalization. 

\subsubsection{Fusion Embedding Centric Contrastive Learning}
To improve the rationality of the fused vector distribution in the latent space and to enforce specific distributional properties such as a spherical distribution, we introduce a fusion embedding centric contrastive learning strategy. In this approach, for each fused vector, a positive sample is created by adding a random Gaussian perturbation $\epsilon$ drawn from a normal distribution with mean zero and variance $\sigma^2$. Negative samples are generated by applying similar perturbations to fused vectors corresponding to other items within the same batch. The learning objective is defined as follows:

\begin{equation}
\label{loss_fus-cl}
\mathcal{L}_{\text{fus-cl}} = -\log \frac{\exp(s(\mathbf{z}_i, \tilde{\mathbf{z}}_i) / \tau)}{\sum_{k=1}^K \exp(s(\mathbf{z}_i, \tilde{\mathbf{z}}_k) / \tau)}
\end{equation}

In this formulation, $s(\cdot, \cdot)$ denotes a similarity function such as cosine similarity, $\tau$ is a temperature parameter that controls the concentration level of the distribution, and $K$ represents the number of negative samples, which are constructed from perturbed fused vectors of other items in the batch.

The proposed contrastive loss encourages each fused vector to remain close to its perturbed positive counterpart, while simultaneously pushing it away from the negatives. This training objective promotes both local consistency and global discriminability in the latent space, thereby ensuring that the fused representations are not only compact within similar instances but also well separated across different items.

Finally, the overall objective function is computed as:
\begin{equation}
\label{all_loss}
\mathcal{L}_{\text{MFL}} = \alpha_1 \mathcal{L}_{\text{title}} + \alpha_2 \mathcal{L}_{\text{cf}} + \alpha_3 \mathcal{L}_{cons} + \alpha_4 \mathcal{L}_{\text{fus-cl}}
\end{equation}

In practice, based on our empirical study, we set $(\alpha_1, \alpha_2, \alpha_3, \alpha_4) = (0.5, 0.25, 0.15, 0.1)$. This multi-objective training enables MFL to learn fusion embeddings that coherently integrate multimodal semantics across genres and demonstrate stronger robustness, better generalization, and improved interpretability.

\subsection{Sparse Attention-based Alignment Layer}

\begin{figure*}[!t]
\centering
\includegraphics[width=6in]{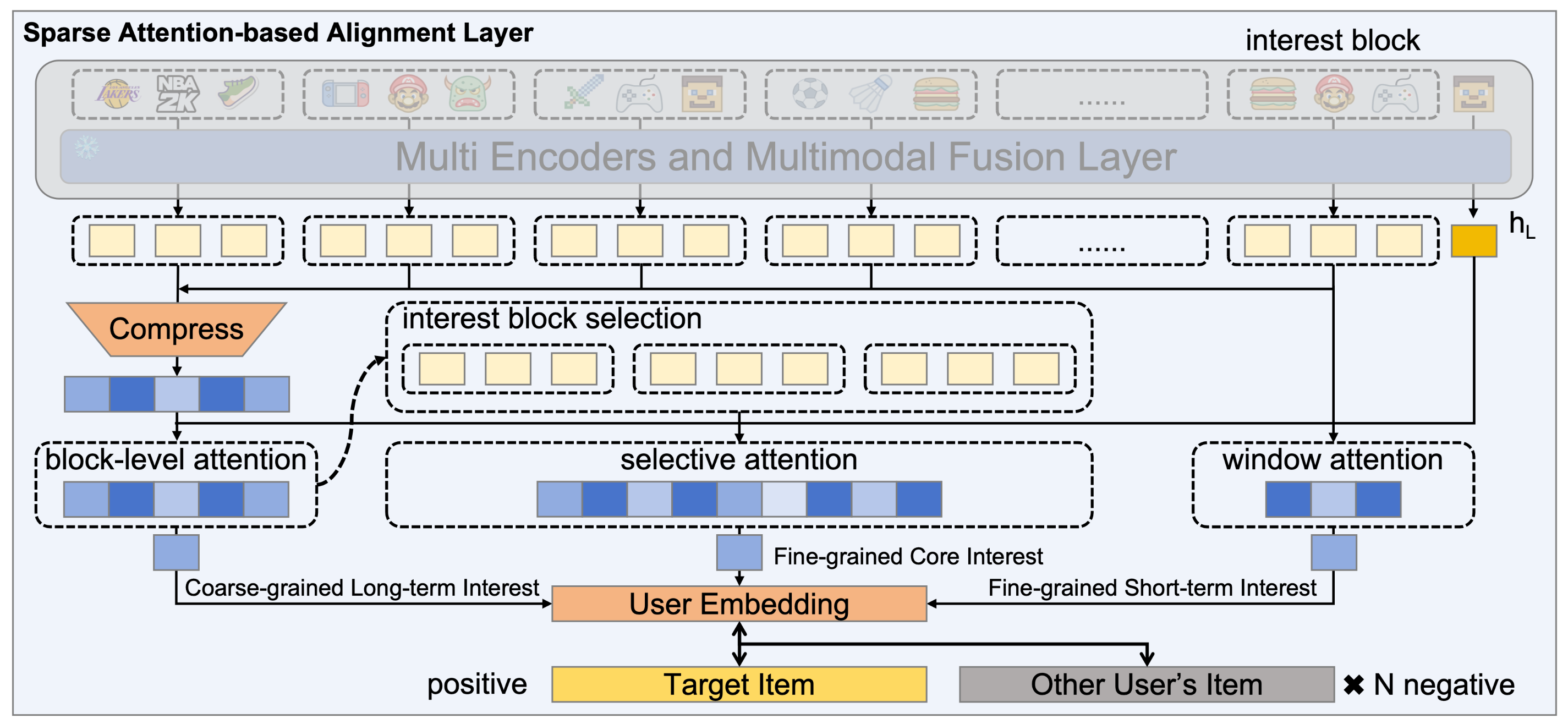}
\Description{Sparse Attention-based Alignment Layer (SAL)}
\caption{Sparse Attention-based Alignment Layer (SAL): The user’s interaction sequence is partitioned into interest blocks. Each item within these blocks is processed by the MFL to acquire its fused multimodal representation. Window Attention extracts Short-term Interest, Block-level Attention extracts coarse-grained Long-term Interest, and Selective Attention identifies the fine-grained core Interest. These attentional outputs are fused into User and Item Embeddings. These embeddings are optimized via contrastive learning, where the pairing of a user and the target item serves as the positive pair, while pairings between the user and items from other users constitute negative pairs.}
\label{fig:fig_2}
\end{figure*}

After the Multimodal Fusion Layer, we obtain an item representation that initially aligns the co-filtering (CF) space and fuses the multimodal information, but this representation still does not fully align with the final semantic space of the recommendation system. To this end, we introduce a Preference Alignment Layer to further map the fused vectors to the recommendation semantic space, and synchronously model user interest preferences to generate user embeddings and final item embeddings.
Unlike previous work that only models user interests at the level of a single item, our proposed Sparse Attention-based Alignment (SAL) module innovatively combines three attention mechanisms to model users' diverse interest preferences from different time scales and interest granularity:

\subsubsection{Fine-grained Short-term Interest Modeling}
The window attention mechanism is introduced to effectively capture users' fine-grained short-term interests and to enhance the model's capability in processing long interaction sequences. The fundamental principle of this mechanism is to selectively attend to a subset of the user's most recent historical interactions, located within a sliding window covering the $W$ items immediately preceding the most recent item. And $W$ means the size of attention window. By doing so, the model can dynamically focus on the user's current preference dynamics, which are often more indicative of immediate behavior than distant historical records.

The size of the attention window is adaptively determined based on the characteristics of each dataset. For instance, in the Mixed-Genre dataset, where users typically exhibit long interaction sequences, a larger window size is employed to capture richer contextual information. In contrast, for datasets such as the Microlens, where user interaction sequences are considerably shorter, a smaller window size is adopted in order to fully assess the effectiveness of the window attention module under limited context.

Specifically, the window size $W$ is set according to the following protocol: when the user's historical sequence length exceeds 30, the window size is set to 8; otherwise, for sequence lengths less than or equal to 30, the window size is set to 4. This strategy ensures that the attention mechanism remains both flexible and sensitive to the underlying data distribution.

Suppose the embedding set of the user's historical interactions is $\mathbf{H}^u = \{ \mathbf{h}_1^u, \mathbf{h}_2^u, \dots, \mathbf{h}_L^u \} \in \mathbb{R}^{L \times d}$, and the current interacted item is $\mathbf{h}_{L}^u$. The window attention is formulated as:

\label{att_short}
\begin{align}
z_{\text{short}} &= \text{Softmax}\left( \frac{\mathbf{Q}_w \mathbf{K}_w^\top}{\sqrt{d}} + \mathbf{M} \right) \mathbf{V}_w \label{loss_z_short} \\
\text{where} \quad 
\mathbf{Q}_w &= \mathbf{h}_{L}^u \mathbf{W}_w^Q, \nonumber \\
\mathbf{K}_w &= \mathbf{H}^u_{[L-W:L]} \mathbf{W}_w^K, \nonumber \\
\mathbf{V}_w &= \mathbf{H}^u_{[L-W:L]} \mathbf{W}_w^V \nonumber
\end{align}

In the above, $\mathbf{W}_w^Q$, $\mathbf{W}_w^K$, and $\mathbf{W}_w^V$ are learnable projection matrices for the query, key, and value representations, respectively. The mask matrix $\mathbf{M} \in \mathbb{R}^{1 \times W}$ ensures that attention is only computed over items within the specified window and is defined as follows:
\begin{equation}
\label{mask_win}
M_{1,j} = \begin{cases}
0 & \text{if } j \in [L-W, L] \\
-\infty & \text{otherwise}
\end{cases}
\end{equation}

This masking operation guarantees that only the most recent $W$ historical interactions contribute to the attention calculation, thereby filtering out outdated information. The resulting short-term interest representation $z_{\text{short}}$ thus encapsulates the user's most current and contextually relevant preferences.

\subsubsection{Coarse-grained Long-term Interest Modeling}

To effectively capture long-term user preferences while mitigating the computational complexity and noise inherent in processing extensive historical sequences, inspired by NSA~\cite{yuan2025native}, we introduce a coarse-grained block-level attention mechanism. This approach addresses two fundamental challenges in long-term interest modeling: the quadratic computational complexity of standard attention mechanisms when applied to lengthy sequences, and the dilution of meaningful signals caused by redundant or noisy interactions scattered throughout the user's history.

The core innovation of our block-level attention lies in its hierarchical abstraction of user interaction patterns. Rather than treating each individual interaction as an atomic unit for attention computation, we aggregate semantically coherent subsequences into interest blocks, which serve as the fundamental units for block-level attention modeling. This design philosophy is motivated by the observation that user interests typically manifest as coherent behavioral patterns spanning multiple consecutive interactions. For instance, a sequence of interactions such as "game reviews, gaming equipment purchases, esports tournament videos" naturally forms a cohesive gaming interest block, while "restaurant reviews, cooking tutorials, recipe searches" constitutes a distinct culinary interest block.

The block-level abstraction offers several theoretical and practical advantages. First, it dramatically reduces the attention matrix dimensionality from $O(L^2)$ to $O(B^2)$, where $L$ represents the original sequence length and $B$ denotes the number of blocks with $B \ll L$. Second, the aggregation process inherently filters out noise from occasional misclicks or atypical interactions, as such outliers become diluted within their respective blocks. Third, this approach naturally captures the temporal evolution of user interests at an appropriate granularity, avoiding both the excessive detail of item-level modeling and the oversimplification of global aggregation.

Formally, we partition the user's historical interaction sequence $\mathbf{H}^u = [\mathbf{h}_1^u, \mathbf{h}_2^u, \ldots, \mathbf{h}_L^u]$ into $B$ consecutive, non-overlapping blocks: $\{ \mathcal{B}_1, \mathcal{B}_2, \ldots, \mathcal{B}_B \}$, where each block $\mathcal{B}_i$ contains a fixed number of consecutive interactions. The representation for each interest block is computed through a learnable aggregation function:

\begin{equation}
\label{block_representation}
\mathbf{b}_i^u = \varphi(\{\mathbf{h}_j^u \mid j \in \mathcal{B}_i\}),
\end{equation}

where $\varphi$ represents a parameterized aggregator that can be implemented as mean pooling or linear projection. In our implementation, we employ a linear projection scheme.

The block-level attention mechanism then computes the relevance of each interest block to the current user context through a standard scaled dot-product attention formulation:

\label{block_att}
\begin{align}
\mathbf{z}_{\text{long}} &= \text{Softmax}\left( \frac{\mathbf{Q}_{\text{b}} \mathbf{K}_{\text{b}}^\top}{\sqrt{d}} \right) \mathbf{V}_{\text{b}} \label{block_attention} \\
\text{where} \quad 
\mathbf{Q}_{\text{b}} &= \mathbf{h}_{L}^u \mathbf{W}_{\text{b}}^Q, \nonumber \\
\mathbf{K}_{\text{b}} &= \mathbf{B}^u \mathbf{W}_{\text{b}}^K, \nonumber \\
\mathbf{V}_{\text{b}} &= \mathbf{B}^u \mathbf{W}_{\text{b}}^V \nonumber
\end{align}

Here, $\mathbf{B}^u = [\mathbf{b}_1^u, \mathbf{b}_2^u, \ldots, \mathbf{b}_B^u] \in \mathbb{R}^{B \times d}$ denotes the matrix of block embeddings, $\mathbf{h}_{L}^u$ represents the current user interacts or query context, and $\mathbf{W}_{\text{b}}^Q, \mathbf{W}_{\text{b}}^K, \mathbf{W}_{\text{b}}^V \in \mathbb{R}^{d \times d}$ are learnable projection matrices for queries, keys, and values respectively.

The resulting attention weights effectively identify which historical interest blocks are most relevant to the user's current context, enabling the model to selectively focus on pertinent long-term preferences while filtering out irrelevant historical patterns. The final output $\mathbf{z}_{\text{long}}$ provides a comprehensive yet focused representation of the user's long-term interests.

\subsubsection{Fine-grained Core Interest Focusing}

While the coarse-grained block-level attention effectively identifies the most relevant interest domains, it inherently sacrifices granular details within each interest block through the aggregation process. To capture the nuanced preferences that distinguish users within the same interest domain, we introduce a fine-grained core attention mechanism that operates on individual items within the most salient interest blocks.

The motivation stems from the observation that users often exhibit highly specific preferences within broader interest categories. For instance, within a travel interest block, one user might demonstrate a strong preference for cultural landmarks, while another might favor outdoor activities. Similarly, in a gaming interest block, preferences might vary significantly between strategy games and action games. These subtle distinctions are crucial for accurate recommendation and cannot be captured through block-level attention alone.

Our fine-grained core attention mechanism addresses this challenge through a two-stage selection process. First, we identify the most contextually relevant interest blocks from the coarse-grained block-level attention output. Second, we apply selective attention exclusively within these selected blocks, thereby maintaining computational efficiency while recovering detailed preference patterns.

Formally, we extract the top-$k$ interest blocks with the highest attention scores from the block-level attention computation. Let $\mathbf{a}_{\text{block}} = \text{Softmax}\left( \frac{\mathbf{Q}_{\text{b}} \mathbf{K}_{\text{b}}^\top}{\sqrt{d}} \right)$ denote the attention weights over blocks. We identify the indices of the $k$ most attended blocks:

\begin{equation}
\label{topK}
\{s_1, s_2, \ldots, s_k\} = \text{TopK}(\mathbf{a}_{\text{block}}, k)
\end{equation}

The corresponding core interest blocks are denoted as 
$\{\mathcal{B}_{s_i}\}_{i=1}^k$. 
For each selected block $\mathcal{B}_{s_i}$, 
we retrieve the original item-level embeddings and concatenate them 
across all selected blocks.:

\begin{equation}
\label{concate_block}
\mathbf{H}_{\text{s}}^u = \text{Concat}(\mathbf{H}_{s_1}^u, \mathbf{H}_{s_2}^u, \ldots, \mathbf{H}_{s_k}^u) \in \mathbb{R}^{N_s \times d}
\end{equation}

where $N_s = \sum_{i=1}^k |\mathcal{B}_{s_i}|$ represents the total number of items across all selected core interest blocks.

The fine-grained selective attention mechanism then computes item-level attention weights within these core interest regions:

\begin{align}
\mathbf{z}_{\text{core}} &= \text{Softmax}\left( \frac{\mathbf{Q}_{\text{s}} \mathbf{K}_{\text{s}}^\top}{\sqrt{d}} \right) \mathbf{V}_{\text{s}} \label{core_attention} \\
\text{where} \quad 
\mathbf{Q}_{\text{s}} &= \mathbf{h}_{L}^u \mathbf{W}_{\text{s}}^Q, \nonumber \\
\mathbf{K}_{\text{s}} &= \mathbf{H}_{\text{s}}^u \mathbf{W}_{\text{s}}^K, \nonumber \\
\mathbf{V}_{\text{s}} &= \mathbf{H}_{\text{s}}^u \mathbf{W}_{\text{s}}^V \nonumber
\end{align}

where $\mathbf{W}_{\text{s}}^Q, \mathbf{W}_{\text{s}}^K, \mathbf{W}_{\text{s}}^V \in \mathbb{R}^{d \times d}$ are learnable projection matrices for the selective attention computation, and $\mathbf{h}_{L}^u$ serves as the query representing the current user context.

This mechanism enables the model to identify specific items within core interest blocks that are most relevant to the current context, effectively recovering detailed preference patterns abstracted during block formation. The computational complexity is significantly lower than applying item-level attention to the entire sequence, as we only compute fine-grained attention over a small subset of the most relevant items, achieving an optimal balance between computational efficiency and representational fidelity.

Finally, a gating mechanism is employed to learn the final user embedding. The user and item embeddings obtained from the three attention modules are optimized through contrastive learning~\cite{yi2019sampling}. Specifically, the embedding of a user and their target item is regarded as a positive pair, while the items of other users are treated as negative pairs.

By modeling user interests in a multi-scale and hierarchical manner, the SAL module can more accurately capture the dynamic changes and diversity of user interests, providing more expressive user and item embeddings for the recommender system.

\section{EXPERIMENTS}
In this section, we detail our experimental design and show performance comparison results, while designing several ablation experiments designed to answer the following questions.  

RQ1: Does our proposed MUFASA have better performance in long sequence recommendation scenarios than other multimodal sequence recommendation models? 

RQ2: Do the different modules of the MUFASA have a positive impact on performance? 

RQ3: Is MUFASA better than other models under different length historical series settings?  
\subsection{Experimental Setup}

\begin{table}[h]
\centering
\caption{Table statistics of the real-world datasets. “Avg” denotes average interaction length; “Sps” indicates interaction sparsity.}
\label{tab:datasets}
\begin{tabular}{l|c|c|c|c|c}
\midrule
Datasets & User & Item & Inter. & Avg & Sps \\ \midrule
Microlens & 100,000 & 19,738 & 719,405 & 7.19 & 99.96\% \\ 
Mixed-Genre & 657,871 & 5,466,601 & 106,290,067  & 161.57 & 99.99\% \\ \midrule
\end{tabular}
\end{table}

Dataset: We used the Mixed-Genre dataset, with each user historical satisfaction sequence greater than 150. The dataset spans multiple content genres including posts, news reports, and micro videos, with each item containing multimodal features: text titles, categories, video frames, audio, and subtitles alongside original item IDs. All user data is anonymized, preserving only user IDs. In addition, we also include MicroLens~\cite{ni2023content}, another multimodal sequential recommendation dataset with text titles, categories, video frames, and audio modalities. Detailed statistics are provided in Table~\ref{tab:datasets}.

Evaluation metrics: For MicroLens, we use a leave-one-out approach to organize our training sets, and use evaluation metrics HitRate@k NDCG@k evaluate the effectiveness of our models. For Mixed-Genre, we use zero-shot recommendation recall@k to evaluate the effect of the model, during the test, we select 200 users who have not appeared in training, each user sorts the historical interaction behavior in chronological order, and takes the last 3 interactions as target items.

We compare MUFASA with other recommendation method: DSSM~\cite{huang2013learning} and LightGCN~\cite{he2020lightgcn}, which only use ID information; YouTube~\cite{covington2016deep} and MMGCN~\cite{woolridge2021sequence}, which directly use multimodal information as side information; and SASRec~\cite{kang2018self}, which uses single-granularity
full-scale attention mechanism.

\subsection{Performance Comparison}

\begin{table}[htbp]
\centering
\caption{Performance comparison on Mircolens. The number following each model name denotes its approximate parameter count, reported in millions of learnable parameters. The best result is denoted in bold, the second best result is denoted with an underline. The “*” denotes the statistical significance ( p < 0.05) of the results of MUFASA compared to the strongest baseline. }
\label{tab:micro_comparison}
\begin{tabular}{c|c|c|c|c}
\midrule
\text{Model} & \text{HR\@10} & \text{HR\@20} & \text{NDCG\@10} & \text{NDCG\@20} \\
\midrule
DSSM-5M     &0.0394 &0.0654 &0.0193 &0.0258 \\
LightGCN-5M &0.0372 &0.0618 &0.0177 &0.0239  \\
YouTube-10M  &0.0392 &\underline{0.0648} &0.0188 &0.0252   \\
MMGCN-10M    &0.0214 &0.0374 &0.0103 &0.0143  \\
SASRec-10M  &\underline{0.0408} &0.0643 &\underline{0.0200} &\underline{0.0259} \\
MUFASA-10M &\textbf{0.0497*} &\textbf{0.0732*} &\textbf{0.0256*} &\textbf{0.0315*} \\
\midrule
\end{tabular}
\end{table}

\begin{table}[htbp]
\centering
\caption{Performance comparison on Mixed-Genre. The “*” denotes the statistical significance ( p < 0.05) of the results of MUFASA compared to the strongest baseline.}
\label{tab:mixed_genre_comparison}
\begin{tabular}{c|c|c|c|c|c}
\midrule
\text{Model} & \text{R\@5} & \text{R\@10} & \text{R\@20} & \text{R\@50} & \text{R\@100} \\
\midrule
SASRec  &0.02 &0.035 &0.07 &0.18 &0.35\\
MUFASA &\textbf{0.045*} &\textbf{0.075*} &\textbf{0.145*} &\textbf{0.32*} &\textbf{0.51*}\\
\midrule
\end{tabular}
\end{table}

\begin{table}[htbp]
\centering
\caption{Performance comparison on Microlens-Long dataset using SASRec as the backbone. Items with fewer than 20 interactions were removed from the original Microlens dataset. SF101, SF50, and MViT refer to models using SlowFast-ResNet101~\cite{feichtenhofer2019slowfast}, SlowFast-ResNet50~\cite{feichtenhofer2019slowfast}, and MViT-B-32x3~\cite{fan2021multiscale} encoders, respectively.}
\label{tab:micro_long_comparison}
\begin{tabular}{c|c|c|c|c|c}
\toprule
 metric     & IDRec  & SF101  & MViT   & SF50 & MUFASA \\
\midrule
HR@10       & 0.1068 & 0.1130 & 0.1178 & 0.1169 & \textbf{0.1262} \\
NDCG@10     & 0.0615 & 0.0606 & 0.0639 & 0.0642 & \textbf{0.0703} \\
\bottomrule
\end{tabular}
\end{table}

The model comparison results are shown in Table \ref{tab:micro_comparison}, \ref{tab:mixed_genre_comparison}, and \ref{tab:micro_long_comparison}. On the Microlens dataset, our model MUFASA achieves a 26.1\% higher HR@10 than DSSM, a 44.7\% higher NDCG@10 than LightGCN, and significantly outperforms YouTube DNN by 12.96\% in HR@20. Compared to the strongest baseline SASRec, MUFASA delivers a 21.8\% absolute improvement in HR@10 and a 28.0\% improvement in NDCG@10, with all gains being statistically significant. For sequences longer than 20 items in Microlens-Long, MUFASA shows still advantages with a 7.13\% higher HR@10 and 9.50\% higher NDCG@10 than previous state-of-the-art models, surpassing all existing baselines. 

On the Mixed-Genre dataset, our model is also better than the previous model. The advantages in R@5 and R@10 evaluation indicators are more obvious, indicating that the relevance of MUFASA recalls is better and more relevant priority recalls. At the same time, the evaluation indexes of R@20, R@50, and R@100 were also significantly improved, and these experimental results proved the effectiveness of MUFASA in multimodal long sequence recommendation scenarios.

\subsection{Online Experiments}
We have deployed MUFASA in our mixed-genre feed product and conducted 10-day online A/B tests across three stages of the recommendation pipeline: the new-item cold-start module, the retrieval module, and the re-ranking module. In each scenario, we allocated 5\% of production traffic to the MUFASA treatment and compared it against the incumbent baseline. In the cold-start module, directly using MUFASA-generated vectors for cold-start retrieval increased the cold-start success rate by 7.24\%. In the retrieval and re-ranking stages, adding MUFASA vectors as auxiliary features to downstream models yielded a cumulative 3.43\% improvement in total media consumption time. These results show that MUFASA can be integrated at multiple points in the production pipeline and deliver consistent commercial gains.

\subsection{Ablation Study}

\begin{table}[htbp]
\centering
\caption{Ablation study on the effect of different modules.}
\label{tab:ablation_study1}
\begin{tabular}{l|c|c|c|c|c}
\hline
\text{Model} & \text{R\@5} & \text{R\@10} & \text{R\@20} & \text{R\@50} & \text{R\@100} \\
\midrule
MUFASA(w/o MFL) &0.01   &0.025 &0.095 &0.205 &0.345\\
MUFASA(w/o SAL) &0.025  &0.075 &0.11  &0.245 &0.46\\
MUFASA &\textbf{0.045}  &\textbf{0.075} &\textbf{0.145} &\textbf{0.32} &\textbf{0.51}\\
\hline
\end{tabular}
\end{table}

\textit{Exploring the effect of different modules.} Table \ref{tab:ablation_study1} presents a detailed ablation analysis on MFL and SAL. This analysis is critical in determining the individual impact of each layer on the performance of model. 

As evidenced in Table \ref{tab:ablation_study1}, removing MFL markedly degrades performance: Recall@100 drops sharply from 0.51 to 0.345, Recall@10 decreases drastically from 0.075 to 0.025 and Recall@5 severely falls from 0.045 to 0.01. These declines highlight MFL’s role in (1)semantic alignment across genres, (2)bridging multimodal features with collaborative filtering embeddings, (3)preventing modality representation collapse and (4)regularizing the fusion space distribution. Without MFL, the model fails to reconcile complementary and conflicting signals across diverse genres and modalities, resulting in semantically confused fusion vectors. Moreover, the absence of MFL causes the fused vectors to deviate from the graph-based collaborative space and ultimately harm recommendation quality.

Table \ref{tab:ablation_study1} demonstrates that SAL layer ablation significantly degrades recommendation performance, evidenced by a critical R@5 recall decline from 0.045 to 0.025. This deterioration originates from SAL’s essential capacity to hierarchically model multi-granular user preferences, integrating both short-term click sequences and long-term interest blocks. SAL achieves this through three key mechanisms: window attention captures short-term user interest; block-level attention identifies crucial interest blocks and captures their inter-block evolutionary patterns, like interest migration between fitness and healthy eating domains; and selection attention models intra-block dependencies within those crucial blocks during continuous topic-coherent viewing. Absent SAL, single-item attention fails to simultaneously represent coherent interest blocks and distinguish fine-grained preferences, critically impairing recommendation task compatibility.

According to the ablation experiment results, both MFL and SAL modules have an irreplaceable role in model performance, and their cascade optimization architecture is the core mechanism to achieve the optimal effect of the model.

\begin{table}[htbp]
\centering
\caption{Analyzing the effect of interaction history length on user behavior.}
\label{tab:ablation_study2}
\begin{tabular}{c|l|c|c|c|c|c}
\midrule
\text{His\_len} & \text{Model} & \text{R@5} & \text{R@10} & \text{R@20} & \text{R@50} & \text{R\@100} \\
\midrule
30  &Variant &0.04  &0.075 &0.115 &0.23 &0.405\\
    &MUFASA    &0.045 &0.075 &0.145 &0.32 &0.51\\
\midrule
150 &Variant &0.035 &0.055 &0.095 &0.275 &0.455\\
    &MUFASA    &0.04  &0.07  &0.13  &0.29  &0.46\\
\midrule
\end{tabular}
\end{table}

\textit{Investigating the impact of the length of user interaction histories.} In the context of sequential recommendation, we aim to evaluate whether sparse attention mechanisms consistently outperform full single-granularity attention mechanisms across user interaction sequences of varying lengths. To this end, we design experimental scenarios with different lengths of user interaction sequences. 

As shown in Table \ref{tab:ablation_study2}, the MUFASA model outperforms its variant with full single-level attention instead of sparse attention across all sequence length settings. These results demonstrate the general effectiveness of the sparse attention mechanism in sequential recommendation tasks: regardless of sequence length, its multi-granularity modeling capability enables more accurate capture of user preference patterns.

\section{Conclusion and Future Work}
This paper proposes MUFASA, a cascade architecture that integrates multimodal and multi-genre alignment and hierarchical interest modeling, including two cascading layers, MFL and SAL. MFL realizes the fusion of text, visual and other modalities, along with the alignment of diverse genres, within a unified semantic space. In addition, MFL provides a preliminary integration of recommendation information into the fused representation via collaborative filtering embeddings. SAL proposes a multi-scale sparse attention mechanism that integrates window attention, block-level atttention and selective attention. This layer explicitly captures the evolution of users' coherent interest blocks, and at the same time identify fine-grained differences within blocks. These modules form a multi-stage training pipeline where MFL generates refined fusion features, subsequently consumed by SAL to construct hierarchical interest preference. The cascaded processing enables coordinated optimization of item and user representations, ultimately achieving state-of-the-art performance as demonstrated in empirical evaluations. This work lays the foundation for multimodal recommendation in semantic alignment and interest modeling, and further research can be carried out on (1)leveraging MLLMs to augment anchor semantics for deeper cross-genre and multi-modal alignment;(2)systematically exploring performance boundaries through parametric scaling of foundation models to further enhance user and item representation.

\bibliographystyle{ACM-Reference-Format}
\bibliography{article}

\appendix

\end{document}